\documentclass[onecolumn, manuscript, 12pt]{aastex631}

\usepackage{graphicx}
\usepackage{xpatch}
\xpatchcmd{\thebibliography}{\twocolumngrid}{}{}{}

\setcitestyle{numbers}

\begin{document}

\title{A Non-Primordial Origin for the Widest Binaries in the Kuiper Belt}

%% Notice placement of commas and superscripts and use of &
%% in the author list

\author{Hunter M. Campbell}
\affiliation{HL Dodge Department of Physics and Astronomy, University of Oklahoma, Norman, OK 73019, USA}
\author{Kalee E. Anderson}
\affiliation{HL Dodge Department of Physics and Astronomy, University of Oklahoma, Norman, OK 73019, USA}
\author{Nathan A. Kaib}
\affiliation{Planetary Science Institute, 1700 E. Fort Lowell, Suite 106, Tucson, AZ 85719, USA}

\begin{abstract}
\bf{Nearly one-third of objects occupying the most circular, coplanar Kuiper belt orbits (the cold classical belt) are binary, and several percent of them are “ultra-wide” binaries (UWBs): $\sim$100-km-sized companions spaced by tens of thousands of km. UWBs are dynamically fragile, and their existence is thought to constrain early Solar System processes and conditions. However, we demonstrate that UWBs can instead attain their wide architectures well after the Solar System’s earliest epochs, when Neptune’s orbital migration implants the modern non-cold, or “dynamic”, Kuiper belt population. During this implantation, cold classical belt binaries are likely to have close encounters with many planetesimals scattered across the region, which can efficiently dissociate any existing UWBs and widen a small fraction of tighter binaries into UWB-like arrangements. Thus, today’s UWBs may not be primordial and cannot be used to constrain the early Solar System as directly as previously surmised.}
\end{abstract}

\section*{Introduction}
At least $\sim$30\% of cold classical belt objects actually consist of binary systems of typically roughly equal-size companions orbiting one another, distinguishing the cold classical belt from the rest of the Kuiper belt where these binaries are significantly rarer \cite{noll08, noll20}. Amongst these binaries, perhaps 1--10\% are UWBs, in which the two components co-orbit with semimajor axes of at least 7\% of the system's Hill radius \cite[though lower UWB fractions are possible since observing biases favor discovery of easily resolved, wide binaries; ][]{lin10, park11, brunzan16}. Due to their extreme companion separations, UWBs are dynamically fragile, and UWB companions can be dissociated through moderately close ($\lesssim1$ Neptunian Hill radius) encounters with Neptune \cite{parkkav10, stonekaib21}, impacts with $\sim$km-scale trans-Neptunian objects (TNOs)\cite{petmous04, parkkav12}, and close ($<1$ system Hill radius) gravitational encounters with other large TNOs\cite{petmous04, campkaib23}. Although they are relatively rare, UWBs' existence and their susceptibility to disruption have been leveraged to constrain their minimum proximity to Neptune in the early solar system\cite{parkkav10} and the number of $\sim$km-sized TNOs residing in the modern Kuiper belt\cite{parkkav12}. In addition, the mutual orbits of TNO binaries have been used to constrain the distribution of angular momentum within planetesimal-forming clumps of pebbles in the early solar system and generating the widest binaries is a challenge for planetesimal formation models\cite{youdgood05, jo07, nes21}. 

A critical implicit assumption upon which these constraints rest is that UWB orbital architectures are in fact primordial. However, past works have shown that UWBs can be derived via the expansion of more tightly bound binaries' semimajor axes through repeated impacts with small TNOs\cite{parkkav12, brunzan16} as well as close, impulse-delivering passages of larger TNOs, since these random impulses drive a diffusion in binary orbital energy \cite{campkaib23}. Nonetheless, significant UWB production in this manner requires numbers of passing/impacting TNO bodies that are well in excess of the modern Kuiper belt's population estimates\cite{parkkav12, campkaib23}. 

This population discrepancy may not invalidate a non-primordial UWB production mechanism, though. While the cold classical belt likely formed in-situ and is stable indefinitely, the same cannot be said for much of the rest of the Kuiper belt\cite{lykmuk05, daw12}. Over time, the orbits of some objects occupying the detached, resonant, scattering, and hot Kuiper belt populations\cite{glad08} (which we group together as a ``dynamic'' Kuiper belt) diffuse until they begin strongly interacting with the giant planets, and they are ultimately ejected from the solar system or trapped into the Oort cloud\cite{oort50, levdun97, dunlev97}. Thus, the population of the dynamic Kuiper belt is smaller today than it has ever been, and its time-averaged value over the solar system's history must be higher than the modern one. 

In addition, it is generally thought that the dynamic Kuiper belt is derived from a primordial belt of planetesimals whose population was $\sim$2--3 orders of magnitude greater than the modern Kuiper belt's\cite{mal95, lev08}. As Neptune migrated away from the Sun and through this primordial belt, some of these belt members become trapped into the modern dynamic Kuiper belt, but this process is extremely inefficient\cite{nes15}. All but 0.1--1\% of primordial belt objects are ejected or trapped in the Oort cloud during this process. However, the removal of a primordial belt object typically takes at least tens of Myrs, during which these eventually-lost objects spend much of their time crossing the cold classical belt. This again suggests that the time-averaged number of TNOs crossing the cold classical belt may be substantially higher than that implied from modern observations. Thus, prior work studying the widening of TNO binaries may have significantly underestimated the effect since they did not consider the larger numbers of TNOs in earlier epochs\cite{campkaib23}.

This is confirmed in Figure 1. Here we analyze the time-varying distribution of orbits within a 4-Gyr numerical simulation of Kuiper belt formation that includes our inner 3 giant planets on their modern orbits while Neptune migrates from 24--30 au\cite{andkaib21}. We use this simulation's orbital distribution to estimate the expected encounter rate between simulation bodies and members of the cold classical Kuiper belt at any given simulation time\cite{campkaib23, pet11, abe21}. In Figure 1A, we plot this encounter rate relative to the encounter rate implied at the end of the simulation (which we assume is analogous to the modern Kuiper belt), and we see that during the simulation's first 35 Myrs, the encounter rate is over 100 times greater than the modern one. Moreover, it remains elevated by an order of magnitude for another 200 Myrs and is at least double the modern encounter rate for $\sim$1 Gyr. Although many of the bodies that transiently cross the cold classical belt do so at higher inclinations, which decreases their individual encounter probabilities, the probabilities are still significant \cite{dell13, abe21}, and the number of such bodies is huge compared to the modern dynamic Kuiper belt. This acts to significantly enhance the number of close encounters we expect a binary to undergo. Figure 1B plots the cumulative encounter flux, and we see that half of all expected encounters should occur during the first 100 Myrs of solar system history, when the Kuiper belt has orbital properties very different from today's. This leads to variations in the encounter velocity distribution, which is shown in Figure 1C. Here we see that the median encounter velocity exceeds 2 km/s for much of the first Gyr, which is well above the 1 km/s encounter velocities assumed in many past works\cite{petmous04, parkkav12, brunzan16}. Thus, considering the long-term evolution of the Kuiper belt is critical when modeling the dynamical evolution of UWBs. 

\begin{figure}
\centering
\includegraphics[scale=1.0]{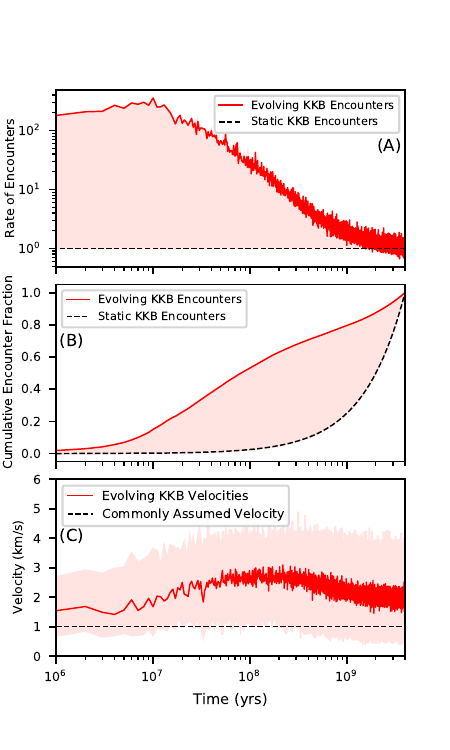}
\caption{{\bf Changes in the TNO encountering population over time.} {\bf A: } The predicted rate of TNO encounters for a cold classical body as a function of time inferred from a simulation of Kuiper belt formation\cite{andkaib21} ({\it red curve}). The predicted rate is plotted and normalized relative to a fixed rate inferred by the end state of the simulation ({\it dashed line}), and the shaded region highlights the excess encounters predicted relative to the assumption of a non-evolving Kuiper belt. {\bf B: } The cumulative flux of TNO encounters for a cold classical body as a function of time inferred from the Kuiper belt formation simulation ({\it red curve}). The cumulative flux is also shown for a static Kuiper belt inferred by the end state of the simulation ({\it dashed curve}), and the difference between the two is shaded. {\bf C: } The median TNO encounter velocity for a cold classical body as a function of time inferred from the Kuiper belt formation simulation ({\it red curve}). The shaded region spans the 5th to 95th encounter velocity percentiles, and the dashed line marks the 1 km/s velocity assumed in prior studies of TNO binary dynamics\cite{parkkav12, brunzan16}.}
\end{figure}

\section*{Results}

To simulate the effects of a time-varying Kuiper belt on TNO binary dynamics, we subject binaries to 4 Gyrs worth of flybys from a diachronic population of TNOs based on Figure 1. Time 0 in all our simulations denotes the completion of  giant planet formation and gas disk dispersal in the outer solar system. Our binaries are given orbits characteristic of a cold classical belt object at 44 au semimajor axis and no eccentricity or inclination. An example of one such simulation is shown in Figure 2a. Here, a TNO binary begins with a semimajor axis (average separation of the companions) that is 3.5\% of its Hill radius, which is half of the minimum separation necessary for UWB classification. After $\sim$500 Myrs of close TNO passages, its mutual semimajor axis has widened to 9\% of its Hill radius, placing it firmly within the UWB regime and quite similar to 2006 JZ$_{81}$, a known UWB\cite{park11}. After another $\sim$500 Myrs, its semimajor axis has been expanded to 15\%, near that of known UWB 2000 CF$_{105}$\cite{noll02}. Following another expansion to a semimajor axis nearly like the widest UWB, 2001 QW$_{322}$\cite{pet08}, additional TNO flybys further expand the binary semimajor axis until it becomes unbound after $\sim$2 Gyrs. Panels 2b and 2c show that these flybys (as well as the solar influence) also drive large changes in the binary's eccentricity and inclination. Thus, Figure 2 documents that a tight TNO binary predicted by past works to be indefinitely stable actually transforms into a UWB within 200 Myrs and is completely dissociated in less than half the solar system's age. 

\begin{figure}
\centering
\includegraphics[scale=1.0]{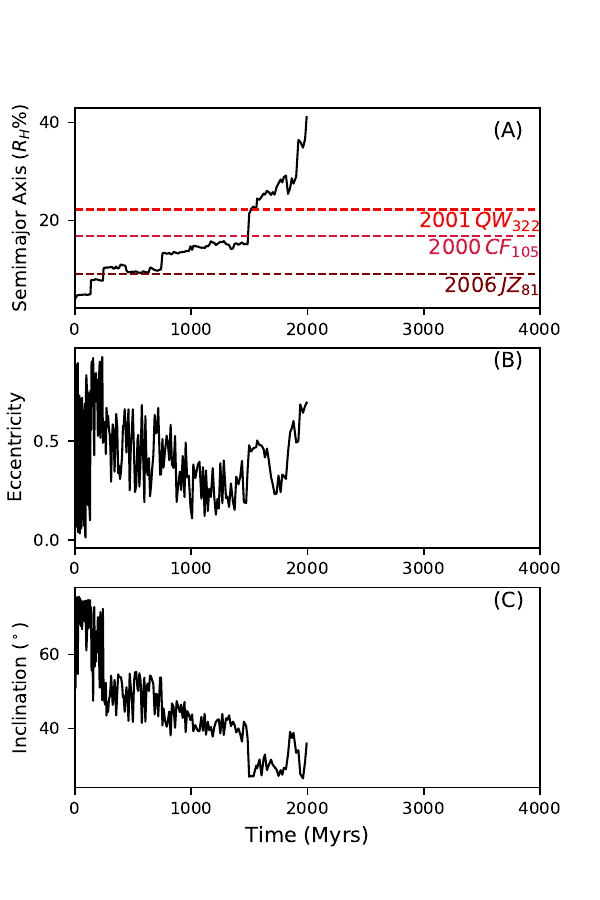}
\caption{{\bf An example of binary widening.} {\bf A: } The evolution of an individual simulated binary's semimajor axis over time as it is exposed to close encounters with TNOs. The semimajor axes of three known UWBs are shown with the labeled dashed lines. {\bf B: } The simulated binary's orbital eccentricity vs time. {\bf C: } The simulated binary's orbital inclination (measured with respect to its heliocentric orbit) as a function of time.}
\end{figure}

We find that the widening of tighter TNO binaries into UWB arrangements is not uncommon in our simulations. The differential velocity impulses our flybys impart on binary companions often act to decrease the gravitational binding of binary orbits and can ultimately unbind them. Such behavior is expected when binaries are subject to gravitational encounters with low-mass bodies passing at velocities much higher than the binary orbital velocity\cite{heg75, bah85}, although semimajor axis decreases are also possible (see section 1.6 of our Supplementary Information). In Figure 3, we plot the distribution of binary semimajor axes sampled during the last Gyr of a simulation in which 1000 binaries with initial semimajor axes of 3--5\% of their Hill radii are subjected to 4 Gyrs of TNO flybys. Here we see that $\sim$9\% of the surviving sample has semimajor axes significantly widen to beyond 0.07 Hill radii, the minimum semimajor axis of the UWB regime\cite{park11,brunzan16}. This percentage depends on the size frequency distribution (SFD) that we assume for the dynamic Kuiper belt, but among observationally favored SFDs\cite{law18}, even the one yielding the weakest set of TNO encounters still causes $\sim$7\% of initially tight binaries to widen into UWBs. (See the section 1.4 of our Supplementary Information for additional considerations of SFDs.) Thus, we should expect TNO passages to widen as many as one tenth of moderately tight (3--5\% Hill radius) binaries into UWBs over the history of the solar system. This does not hold true, however, for even tighter binaries, as systems with initial mutual semimajor axes of 1--2\% Hill radius virtually never widen to UWB status (see section 1.6 of our Supplementary Information). 

\begin{figure}
\centering
\includegraphics[scale=1.0]{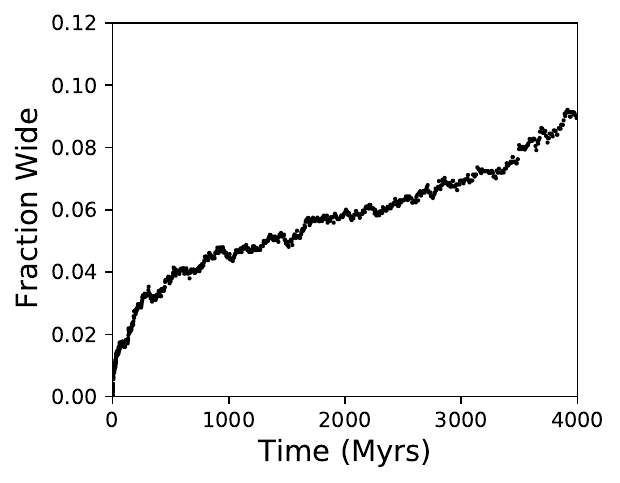}
\caption{{\bf Binary widening efficiency.} For 1000 simulated binaries with initial semimajor axes between 3--5\% of Hill radius, the fraction of bound systems that evolve to semimajor axes over 7\% of Hill radius is shown as a function of time.}
\end{figure}

Of course, our simulated UWBs generated from widened tight binaries have a distribution of orbits, and this can be compared with the orbits of observed UWBs. In Figure 4a, we compare our simulated UWB semimajor axes with those from the 9 known UWB systems\cite{park11, grun19}. We see that the distributions look qualitatively similar, and a Kolmogorov-Smirnov test comparing the two distributions returns a $p$-value of 0.09, indicating that we cannot reject the null hypothesis that the two distributions sample the same underlying distribution with 2-$\sigma$ confidence. 

We can similarly compare the eccentricity distributions of simulated and observed UWBs. In this case, our initial conditions are designed to explicitly test whether observed UWBs can be derived from tighter binaries, as our initial binary eccentricity distribution is simply the observed eccentricities for non-UWB binaries\cite{grun19}. It is well known that UWBs have hotter eccentricities than non-UWBs but that they are also cooler than a thermalized distribution ($<e>\simeq0.7$)\cite{park11,grun19}. In Figure 4b, we show the eccentricity distribution of UWBs in our simulations that are generated from the gradual widening of tighter binary systems. We see that the simulated and observed distributions appear similar, and a K-S test returns a $p$-value of 0.94, indicating that the statistical differences in these two distributions are not significant enough to rule out the generation of UWB eccentricities through the widening of tighter, less eccentric binaries. 

We also compare the inclinations of simulated and observed UWBs in Figure 4c. Once again, it has been posited that observed UWBs have a different inclination distribution than tighter binaries, with UWBs exhibiting a greater paucity of polar orbits\cite{park11, grun19}. With a sample size of just 9 observed UWBs, it is not clear that this observed difference is statistically significant, as a K-S test comparing the two observed distributions yields a $p$-value of 0.28. Nevertheless, if we begin with 1000 tight binaries with the observed tighter inclination distribution, Figure 4c shows that the simulated UWBs that we generate possess fewer polar orbits than the tighter binaries. This appears to be an effect of the Kozai-Lidov mechanism\cite{koz62, lid62}. Kozai-Lidov cycles drive binaries to large eccentricities (and large apocenter) at the minimum inclination of the cycle. Thus, Kozai-cycling binaries have larger average separations at their low inclinations, which allows TNO passages to deliver stronger impulses during this phase, enhancing the probability that UWBs will be generated at such inclinations. Although the decrease in polar orbits is significantly less dramatic among our simulated UWBs, we again stress that the observed UWB sample consists of just 9 systems, and a K-S test comparing the simulated and observed UWBs cannot reject the null hypothesis with 2-$\sigma$ confidence ($p=0.34$). 

\begin{figure}
\centering
\includegraphics[scale=1.0]{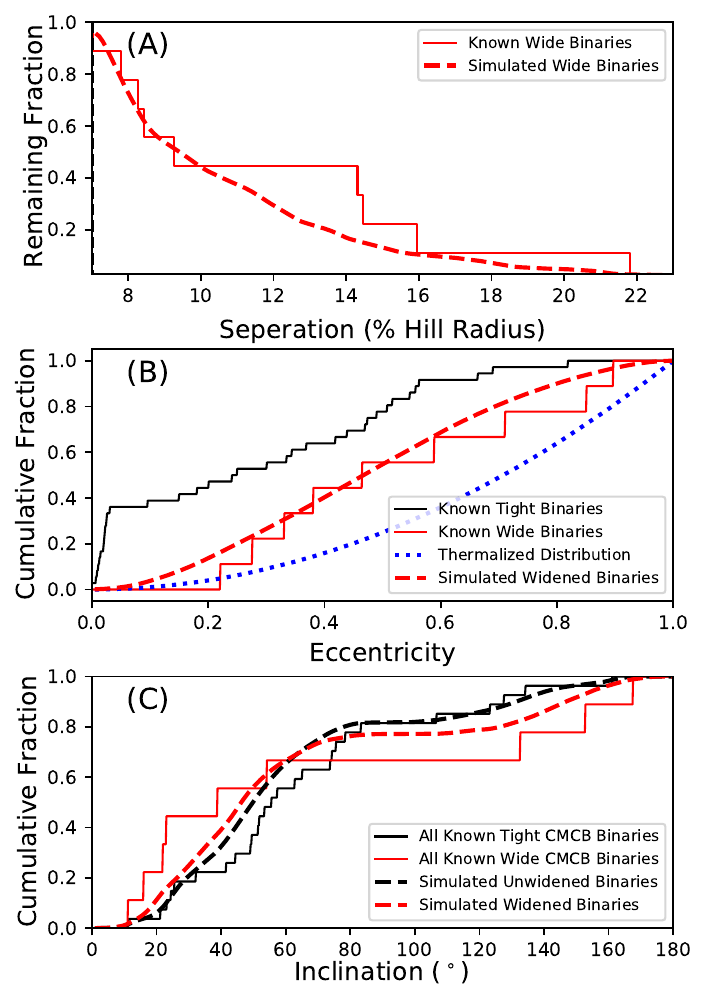}
\caption{{\bf Orbital distributions of widened binaries after 4 Gyrs.} {\bf A: } The distribution of the semimajor axes of simulated ({\it dashed}) UWBs formed through the widening of tight binaries  is compared against the observed ({\it solid}) distribution of UWB semimajor axes. {\bf B: } The distribution of the eccentricities of simulated ({\it dashed}) UWBs formed through the widening of tight binaries is compared against the observed ({\it solid}) distribution of UWB eccentricities. A thermal eccentricity distribution is also shown ({\it dotted}). {\bf C: } The distribution of the binary inclinations of simulated ({\it dashed red}) UWBs formed through the widening of tight binaries is compared against the observed ({\it solid red}) distribution of UWB eccentricities in the cold classical belt (CMCB). The simulated ({\it dashed black}) and observed distributions ({\it solid black}) for non-UWB binaries are also shown.}
\end{figure}

Finally, we can also use our simulations to study the survival of hypothetically primordial UWBs throughout the entire evolution of the Kuiper belt. To do this, we assemble 1000 binaries with initial system masses and semimajor axes matching 2001 QW$_{322}$ and subject them to 4 Gyrs worth of TNO passages. The surviving fraction as a function of time is shown in Figure 5. Here we see that the large majority are dissociated over the history of the solar system, as only 5.1\% reach the end of the simulation. However, despite most binary evolution being toward increased separation, the vast majority of surviving binaries here persist by decreasing in separation. Smaller semimajor axes help shield them from TNO passage effects, and most such binaries reach the end of the simulation because they evolve to a smaller semimajor axis ($<0.8$ of their original value, roughly the same separation as 2000 CF$_{105}$, the next widest binary). Only 1.7\% of our binaries finish with semimajor axes that are greater than or equal to 0.8 of the semimajor axis of 2001 QW$_{322}$ (0.22 Hill radius). The rate of destruction is again dependent on our assumed dynamic SFD, but among observationally favored SFDs\cite{law18}, our weakest encounter sets still yield just 2.5\% of binaries with final semimajor axes beyond 80\% of 2001 QW$_{322}$. This implies that if the binary orbit of 2001 QW$_{322}$ is indeed primordial, it is the remnant of a primordial population that is $\sim$40--60 times greater than today's. 

\begin{figure}
\centering
\includegraphics[scale=1.0]{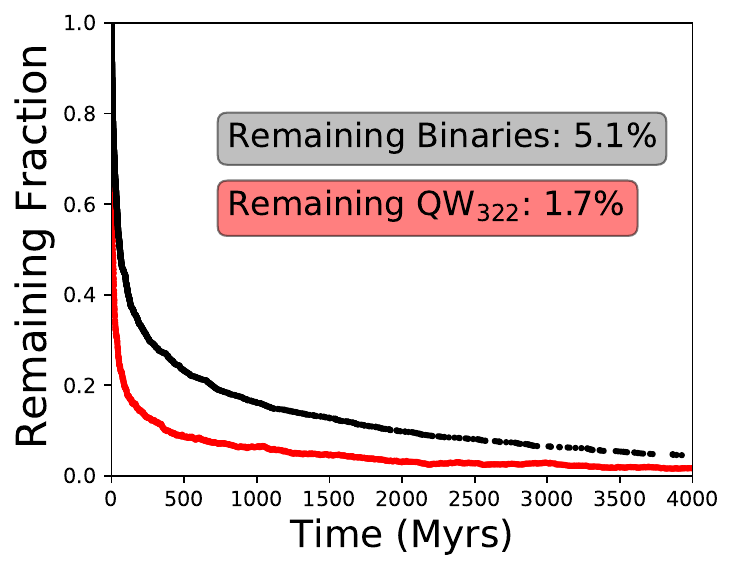}
\caption{{\bf Destruction of primordial UWBs.} The fraction of simulated 2001 QW$_{322}$ binaries that remain as a function of time. The black curve represents all bound binaries, and the red curve represents only binaries with a semimajor axis at least 80\% of that observed for 2001 QW$_{322}$.}
\end{figure}

\section*{Discussion}

Most prior work has underestimated the orbital evolution and instability of TNO binaries. This is for two reasons. The first is that more attention and modeling has been devoted to considering the effects of velocity impulses from collisions with small TNOs rather than close encounters with TNOs. While collisions can indeed alter the orbital evolution of TNO binaries, the magnitude of a collisional impulse does not depend on the binary orbital architecture. This is not true for TNO close encounters, in which the perturbation effectively results from a tidal force, and binaries with larger separations experience larger perturbations. Thus, a sequence of close encounters on a binary has a potential for a stronger multiplier effect, wherein the effects of later encounters are more dynamically consequential for the binary evolution if early encounters result in some degree of binary widening. 

The second reason is that the total population of the dynamic Kuiper belt has decreased substantially as the solar system has aged. Objects within the cold classical belt experienced close TNO encounters at a much higher rate in early solar system epochs compared to the modern epoch. Thus, the total number of encounters that TNO binaries have suffered is substantially higher than that implied by a static backwards projection of modern Kuiper belt conditions. 

Our simulations show that if the widest UWB orbits are in fact primordial, then they are the remnants of a much larger primordial binary population that has largely been dissociated via impulses from TNO passages. Because these dissociated members would be a major contributor to today's population of single objects within the cold classical belt, this primordial origin scenario may be in conflict with the observation that cold classical singles have a distinctly redder color distribution than cold classical binaries \cite{fras21}. In particular, of the 83 cold classical singles with accurate color measurements, just 2 (2.4\%) have spectral slopes below 17\%, while 8 of 30 (27\%) cold classical binaries have such flat slopes. These notably flat-sloped binaries include 2001 QW$_{322}$, which should now accompany many flat-sloped single counterparts if the binary is a rare primordial survivor of TNO passages. \cite[It would have been subject to destructive TNO passages whether Neptune dynamically pushed it out into the cold belt or if it always resided in the cold belt;][]{fras17}. It is not clear how this stark discrepancy between binary and single colors could be maintained if such a large primordial UWB population existed and was then subsequently mostly converted into singles. 

Our work shows that close TNO passages will expand the orbits of some tighter binaries to generate a population of UWBs with an efficiency and orbital distribution that may explain the observed population of UWBs. Thus, it is possible that the orbits of known UWBs are not primordial and that most of these systems only attained their dynamically fragile states at least hundreds of Myrs after the solar system's formation. If this is the case, then UWB orbits do not necessarily constrain the degree of their minimum proximity to Neptune, the number of small impacting bodies in the early outer solar system, or the angular momenta of planetesimal formation models. 

\subsection*{Methods}
Our simulations of TNO binary evolution use a modified version of the SWIFT RMVS4 integrator, chosen for its ability to accurately integrate test particles through very high eccentricity \cite{levdun94}. We simulate the widening of 1000 tight (3--5\% Hill radius) binaries. Hill radius here is the gravitational influence of a system orbiting at 44 au, defined identically to similar prior work \cite{parkkav12, campkaib23}. The lower bound of this separation range is chosen because there is diminishing semimajor axis evolution below this (see section 1.6 of our Supplementary Information), and the upper bound is chosen because it requires significant semimajor axis evolution ($\sim$50\% growth) for binaries to reach the UWB regime. In addition, we also simulate the dynamical survival of 1000 binaries resembling 2001 QW$_{322}$. Simulations are run in batches of 10 massless test particles placed in orbit about a common central mass that possesses the combined mass of 2001 QW$_{322}$, or $\sim$$2.1\times10^{18}$ kg\cite{park11}. (See section 1.5 of our Supplementary Information for consideration of a lower system mass.) Binary (test particle) inclinations and eccentricities are randomly drawn from the distributions observed among non-ultra-wide binaries \cite[whose inclinations resemble the predictions of recent streaming instability simulations;][]{Nevorn19, grun19}, and arguments of pericenter, longitudes of ascending node, and mean anomalies are all randomly drawn from uniform distributions. Simulations of tight binaries employ a 16-day timestep (1/20 of a 3\% Hill radius orbital period), and simulations of 2001 QW$_{322}$ use a 50-day timestep\cite{campkaib23}. 

In addition to including the Sun on a fixed circular orbit at 44 au, each batch of 10 binary systems is perturbed by a unique set of passing TNOs. The rate and velocity of such passages are determined from the 1-Myr outputs of a ``grainy slow'' simulation of Kuiper belt formation that excludes a hypothetical undiscovered planet and includes the inner 3 giant planets on their modern orbits as well as Neptune migrating to 30 au through a disk of test particles\cite{andkaib21}. \cite[An alternative ``smooth fast'' formation model is considered in section 1.3 of our Supplementary Information;][]{kaibshep16}. During each 1-Myr output, the encounter probability between the dynamic bodies of the Kuiper belt formation simulation and the observed cold classical belt as well as the cold classical belt self-interaction \cite{pet11, abe21} is numerically computed, in addition to the encounter speeds\cite{campkaib23}. The total number of dynamic bodies at each Kuiper belt formation simulation step is calculated assuming that there are 250,000 dynamic bodies with $H_r < 8.66$ at $t=4$ Gyrs in our Kuiper belt formation simulation, which matches observations\cite{abe21}, and then scaling accordingly. This population number is then extrapolated down to radii of 20 km using a magnitude frequency distribution (MFD) and an assumed albedo of 0.04\cite{glad01, law18}. The combined influence of passing bodies of smaller radius than 20 km is far less significant than the combined influence of ones larger\cite{campkaib23}. The MFDs we test have a faint-end power law index of 0.4 and a bright-end index of 0.9. We run all of our simulations twice, differing in the point of transition between the two MFD power laws (``the knee''): once at $r=50$ km and again at $r=85$ km, which are both allowed by observations\cite{law18}. The MFD of our cold classical belt is an exponentially tapered power law \cite{kav21} and its population size is similarly scaled over time. Finally, we assume a bulk density of 1 g/cm$^3$ to convert passing TNO radii to masses\cite{parkkav12, brunzan16}. In this manner, 4 Gyrs of TNO encounters are generated for each batch of simulated binaries, but the inter-encounter time is shortened to increase simulation throughput\cite{campkaib23}. Potential collisions with smaller TNOs are notably absent from our simulations. This is done to study and isolate the role of passages, but inclusion of collisions could likely enhance the dynamical behaviors we document here \cite{brunzan16}.

\subsection*{Data Availability}
The simulation outputs necessary to replicate all figures and analyses in this work will be provided on the publicly accessible repository, Harvard Dataverse (https://doi.org/10.7910/DVN/LI1NJF). 

\subsection*{Code Availability}
The analysis codes necessary to replicate all figures and analyses in this work will be provided on the publicly accessible repository, Harvard Dataverse (https://doi.org/10.7910/DVN/LI1NJF). Our numerical simulation source code is available at https://github.com/nathankaib/Swift\_KBOBin. 

\subsection*{Acknowledgements}
This work was supported by NASA Emerging Worlds grants No. 80NSSC18K0600 and No. 80NSSC23K0868. Our computing was performed at the OU Supercomputing Center for Education \& Research (OSCER) at the University of Oklahoma (OU). 

\subsection*{Author Contributions}
HMC modified and tested the numerical binary code and oversaw the running of the simulations. KEA extracted the necessary data from the Kuiper belt formation simulations. All authors contributed with the interpretation and discussion of the results, writing and editing of the manuscript. 

\subsection*{Competing interests}
The authors declare that they have no competing financial interests.

\subsection*{Correspondence}
Correspondence and requests for materials should be addressed to Hunter Campbell (Hunter.M.Campbell-1@ou.edu).

\clearpage
\renewcommand{\figurename}{Supplementary Figure}
\setcounter{figure}{0}

\section{Supplementary Information}

\subsection{Timestep Considerations}

In the absence of TNO passages, a timestep that is 1/10 to 1/20 of our binaries' orbital periods should suffice for accurate integrations. However, the timescales for encounters (impact parameter over velocity) can sometimes be much shorter than this. To handle the fastest encounters, we employ the Impulse Approximation, since the orbital velocities are typically much slower than the TNO passage velocities. However, we wish to directly integrate slower encounters when the Impulse Approximation may be less accurate. 

It is not clear when to switch from Impulse Approximation implementation to direct integration, so we perform a set of numerical experiments to guide our choice. This set is a repeat of the main paper's simulations modeling the survival of 2001 QW$_{322}$. The only difference is that we vary the encounter timescale, above which we switch from Impulse Approximation to direct integration. Our integration step size for these simulations is 50 days, and our encounter timescale switchover is varied from 2.5 days to 500 days. As we can see in Table 1, for switchover timescales between 10--500 days, the survival rate of 2001 QW$_{322}$ analogs is between 5.0--5.6\% (it is 5.1\% in the main paper's simulations) with no obvious trend in survival rates across these switchover timescales. Below 10 days, however, the survival rate falls steeply. Thus, we conclude that encounter timescale switchovers above 10 days are sufficiently accurate, and we choose 50 days for the main paper simulations since it is well above this value.

\begin{table}
\centering
\begin{tabular}{ |c|c|  }
\hline
Impulse Approximation Cutoff & Survival Rate \\
\hline
\hline
2.5 days& 0.8\%\\
5 days& 3.0\%\\
10 days& 5.2\%\\
20 days& 5.0\%\\
50 days& 5.1\%\\
100 days& 5.4\%\\
300 days& 5.0\%\\
500 days& 5.6\%\\
\hline
\end{tabular}
\caption{{\bf Survival Rate of 2001 QW$_{322}$ as a function of our impulse approximation cutoff.} The survival rates of 2001 QW$_{322}$ are compared assuming a different encounter timescale trigger for using the Impulse Approximation over direct integration. For the entirety of all other 2001 QW$_{322}$ simulations, a cutoff time of 50 days is employed.}
\end{table} 

\subsection{UWB Survival Under Different Passing TNO Fluxes}

Because the perturbing TNO flux that a classical binary will experience may not be completely constrained, we also test the survival likelihood of 2001 QW$_{322}$ with modifiers on the net flux of TNO flybys that occur. These 2001 QW$_{322}$ survival simulations are identical to those presented in the main paper except that the TNO flux is scaled up or scaled down (with the SFD held fixed). The results of these simulations are shown in Supplementary Figure 1. As expected, increasing the flux only lowers 2001 QW$_{322}$'s chance of survival. The survival probability of 2001 QW$_{322}$ falls to $\sim$2\% if the flux of passing TNOs is doubled, and it only exceeds 10\% when the flux is roughly halved. The Kuiper belt flux we assume would have to be reduced to 1/5 of the value in our main paper simulations for 2001 QW$_{322}$ to survive 50\% of the time.  

\begin{figure}
\centering
\includegraphics[scale=1.0]{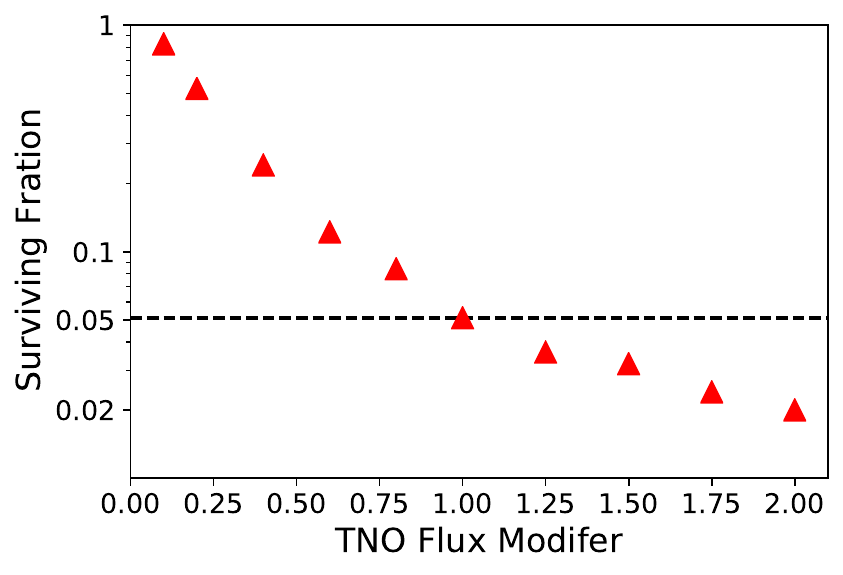}
\caption{{\bf The survival likelihood of 2001 QW$_{322}$ as a function of the overall flux of perturbing TNOs.} A TNO Flux modifier of 1 represents our unaltered 2001 QW$_{322}$ simulations as shown in Figure 5. This survival likelihood is also marked by the horizontal dotted line.}
\end{figure}

\subsection{TNO Binary Evolution in Alternate Kuiper Belt Formation Model}

To explore how our results vary with different Kuiper belt formation models, we rerun two of our simulations with TNO passages expected from an alternate model of Kuiper belt formation. This alternate Kuiper belt formation model assumes a smoother migration for Neptune's semimajor axis and a migration timescale that is three times shorter than the model used in the main paper \cite{kaibshep16}. Just as in the main paper, we sample the test particle orbital distribution every 1 Myrs and build a set of TNO passages based on the total population, encounter probability, and encounter velocities of this sampled orbital distribution. Just as in the main paper, our TNO passage rates are set by assuming that the end state of the Kuiper belt formation simulation possesses as many bodies as is observationally estimated in the actual Kuiper belt. 

As depicted in Supplementary Figure 2, this alternate Kuiper belt migration model leads to a significant net decrease in the survival likelihood of 2001 QW$_{322}$ compared to the main paper simulations. Although the total fraction of 2001 QW$_{322}$ analogs that survive for 4 Gyrs falls by about 40\% (from 5.1\% to 3.1\%) the falloff is steeper for analogs that retain separations like 2001 QW$_{322}$ at the end of simulation. This falls from 1.7\% in the main paper to 0.5\%, or by a factor of $\sim$3.  

\begin{figure}
\centering
\includegraphics[scale=1.0]{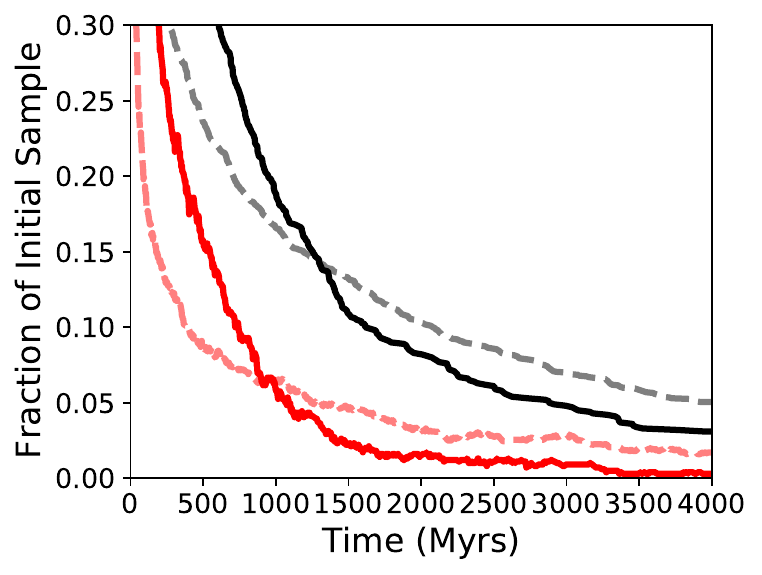}
\caption{{\bf Loss rate of 2001 QW$_{322}$ with an alternate Kuiper belt formation model.} The black curve represents all remaining binaries while the red curve represents all remaining binaries which retain a semimajor axis at least 80\% of that observed for 2001 QW$_{322}$. Here, an alternate Kuiper belt formation model is used in our simulations. Compare to Figure 5 of the main paper.}
\end{figure}  

When this alternate model is applied to initially tight binaries, the final resulting eccentricity and inclination distributions are not significantly different from the main paper. These are shown in Supplementary Figure 3, and, again, we cannot use a K-S test to confidently reject the null hypothesis that the synthetic distributions and their corresponding observed ones do not share the same underlying distribution. In this alternate model, we also see a heightened UWB production from initially tighter binaries. Now $\sim$11\% of surviving binaries have widened past the 7\% ultra-wide threshold, a modest increase from the 9\% we see in the main paper. Thus, for two different Kuiper belt formation models that replicate major features of the observed modern belt, we find that the survival of a primordial 2001 QW$_{322}$ is very unlikely, and we find that a significant fraction of tighter binaries can be widened to UWBs over 4 Gyrs. 

\begin{figure}
\centering
\includegraphics[scale=1.0]{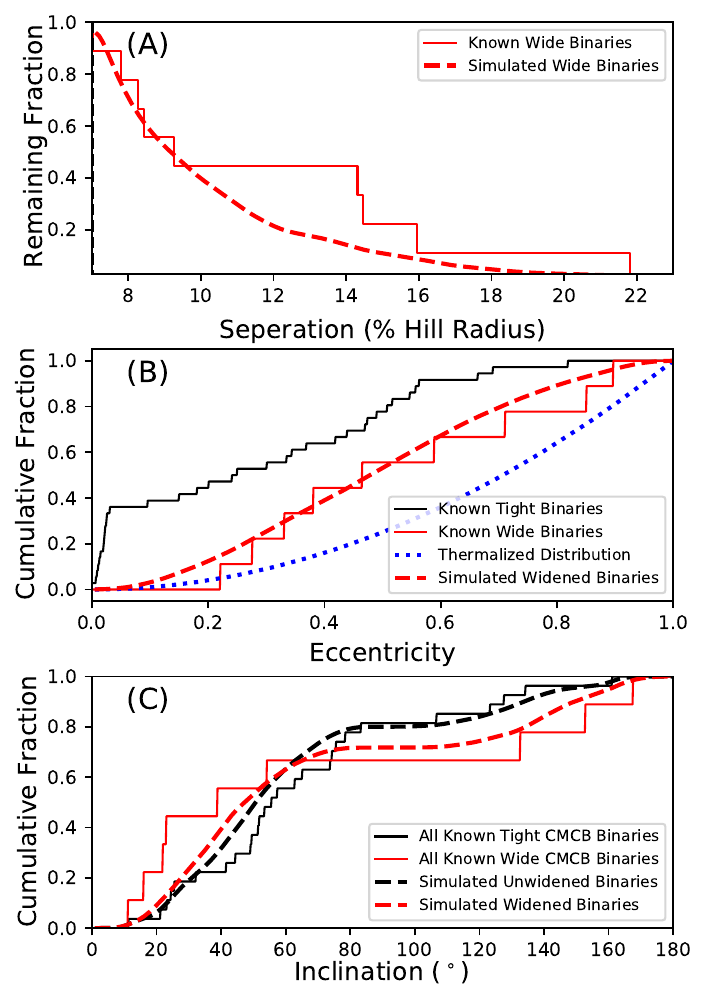}
\caption{{\bf Orbital distributions of widened binaries with an alternate KB formation model. } {\bf A: } The semimajor axis distribution of widened binaries compared to the observed UWB distribution. {\bf B: } The eccentricity distribution of widened binaries compared to the known distribution of tight binaries, wide binaries, and a thermalized distribution. {\bf C: } The inclination distribution of simulated widened and tight binaries compared to the observed distributions of wide and tight binaries. Compare to Figure 4 of the main paper. Here, an alternate Kuiper belt formation model is employed in all simulations.}
\end{figure} 

\subsection{Alternate Size Frequency Distributions}

In an additional series of simulations, we study the sensitivity of the main paper's results to our chosen SFD. In the main paper, we assume that passing bodies from the dynamic Kuiper belt follow a size frequency distribution comprised of two power-laws: a shallow faint-end distribution transitioning to a steeper bright-end distribution at a break radius of 85 km. However, it has recently been proposed that the dynamic Kuiper belt could possess an SFD that is identical to the cold classical belt's recently proposed exponentially tapered distribution \citep{kav21, petit23}. 

We therefore run a new set of simulations modeling the survival of 2001 QW$_{322}$ analogs in which the dynamic Kuiper belt takes the same SFD (exponentially tapered) as the cold classical belt. Our results are shown in Supplementary Figure 4. The results look very similar to those in the main paper, but with a slightly lower survival probability. In this new set of simulations, 4.8\% of binaries survive for 4 Gyrs, and 1.3\% finish the simulation with separations similar to 2001 QW$_{322}$.  

\begin{figure}
\centering
\includegraphics[scale=1.0]{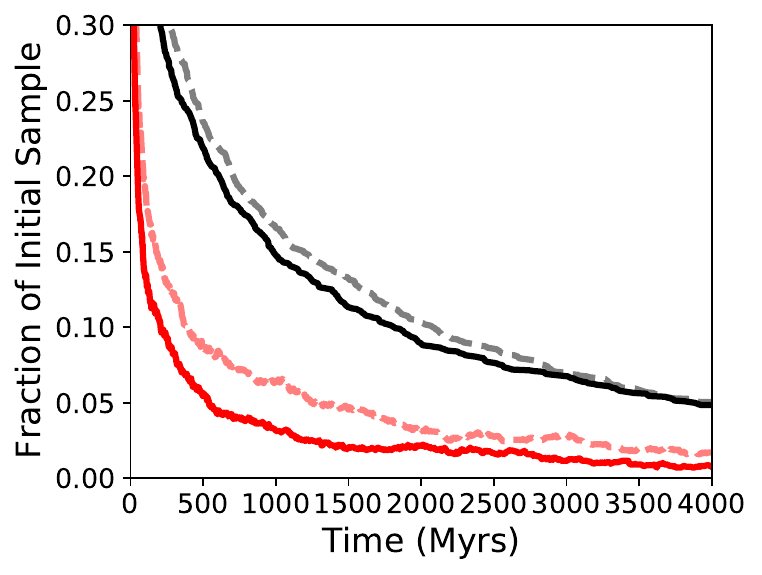}
\caption{{\bf Loss rate of 2001 QW$_{322}$ with an alternate dynamic Kuiper belt size frequency distribution.} The black curve represents all remaining binaries while the red curve represents all remaining binaries which retain a semimajor axis at least 80\% of that observed for 2001 QW$_{322}$. Compare to Figure 5 of the main paper. Here, the dynamic Kuiper belt size distribution is identical to that of the CKB.}
\end{figure} 

We also run an additional set of simulations to study the role that large TNOs (approaching dwarf planet radii) play in the evolution of our binaries. In this new set of simulations, we clip our SFD at a radius of 500 km so that no passing bodies are larger than this radius. In this clipped-SFD simulation, we again study the dynamical survival of 2001 QW$_{322}$ analogs, and the results are shown in Supplementary Figure 5. Here we see nearly identical results to the main paper. 5.8\% of our analogs survive for 4 Gyrs, and 1.6\% finish the simulations with separations similar to 2001 QW$_{322}$. We therefore find that the dynamical evolution of TNO binaries documented in the main paper is largely preserved if we employ other plausible SFDs and is also insensitive to our assumptions about the population of TNOs with radii over 500 km. 

\begin{figure}
\centering
\includegraphics[scale=1.0]{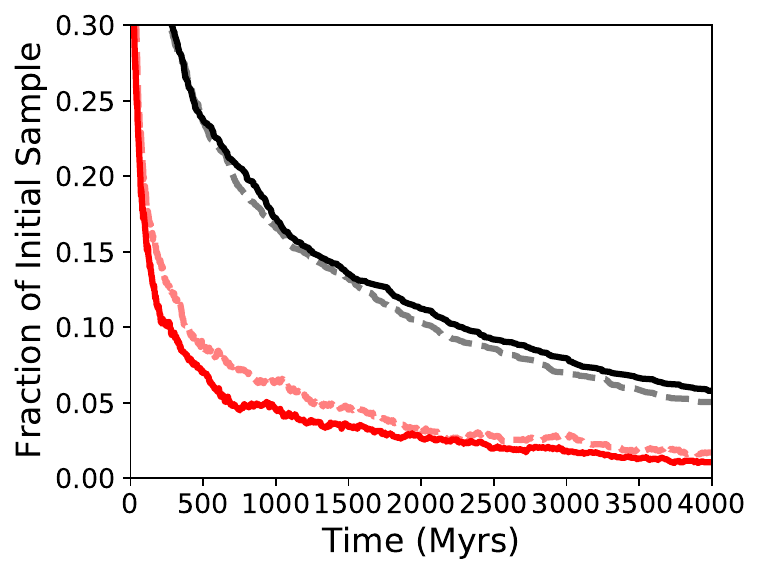}
\caption{{\bf Loss rate of 2001 QW$_{322}$ with an impactor size cutoff.} The black curve represents all remaining binaries while the red curve represents all remaining binaries which retain a semimajor axis at least 80\% of that observed for 2001 QW$_{322}$. Compare to Figure 5 of the main paper. Here, the maximum size of any perturbing TNO is 500 km, all larger potential encounters are discarded.}
\end{figure}  

\subsection{Lower Mass Widening Binaries}

All of our simulations that model the evolution of initially less-than-ultra-wide binaries are done with a binary mass set to that of 2001 QW$_{322}$. However, because 2001 QW$_{322}$ is the most massive ultra-wide binary yet observed, we seek to determine how such binaries might widen if they have less mass. Spanning the range of ultra-wide masses known, we run a new series of initially less wide binaries with the mass of 2000 CF$_{105}$, the least massive ultra-wide binary yet known and over an order of magnitude less massive than 2001 QW$_{322}$, or $\sim$$1.9\times10^{17}$ kg.

This difference in mass does not cause significant changes in the final orbital distribution of the binaries, as shown in Supplementary Figure 6 and in comparison to Figure 4 of the main paper. We see that the semimajor axis and eccentricity distributions of the ultra-wide binaries generated in our simulations compare well with the observed ultra-wide distributions. Again, just as in the main paper, the simulated ultra-wide inclination distribution is slightly more planar than the distribution of simulated tighter binaries. One notable difference between the simulations modeling 2001 QW$_{322}$ masses and 2000 CF$_{105}$ masses, however, is the resulting fraction of widened binaries. For initially moderately tight (3--5\% Hill radii) with 2000 CF$_{105}$ masses, the fraction of surviving binaries that become ultra-wide after 4 Gyrs is 6\%, compared to the 9\% seen in the same initial binary semimajor axes with 2001 QW$_{322}$ masses. 

\begin{figure}
\centering
\includegraphics[scale=0.8]{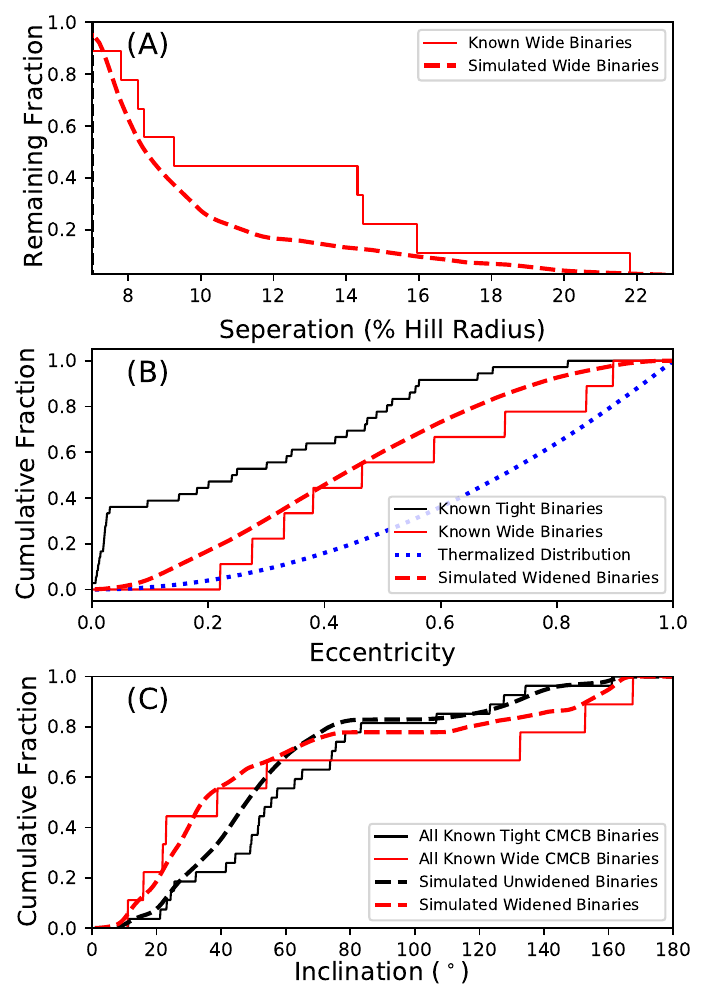}
\caption{{\bf Final orbital properties of initially tight binaries of lower mass.} {\bf A: } The semimajor axis distribution of widened binaries compared to the observed UWB distribution. {\bf B: } The eccentricity distribution of widened binaries compared to the known distribution of tight binaries, wide binaries, and a thermal distribution. {\bf C: } The inclination distribution of simulated widened and tight binaries compared to the observed distributions of wide and tight binaries. Compare to Figure 4 of the main paper. Here, a lower binary mass (that of 2001 CF$_{105}$) is used.}
\end{figure}

\subsection{Dynamical Inertness of Still Tighter Binaries}

While we see significant portions of initially tight binaries become ultra-wide over 4 billion years, there must be a point where initial semimajor axes are not large enough to enable this widening. To explore this, we also integrate an ensemble of 1000 binaries with semimajor axes between 1--2\% of their Hill radii for 4 Gyrs. In contrast to our initially 3--5\% Hill radius semimajor axis binaries, this still tighter population does not generate ultra-wide binaries. 

Supplementary Figure 7 shows the final semimajor axis distribution of these binaries in terms of percentage change relative to their initial semimajor axes. We see that virtually no binaries experience more than a doubling of their semimajor axes, putting an upper limit of 4\% Hill radius on the final ensemble. In contrast, our binaries beginning with initial semimajor axes of 3--5\% Hill radius can undergo widening of over 150\% of their initial separations, or 0.21 Hill radius. The widening is even more dramatic for our 2001 QW$_{322}$ analogs. Among these binaries, $\sim$1/3 of our ensemble undergoes at least a doubling of their semimajor axes (often on their way to complete dissociation), and $\sim$15\% experience widening of over 150\% of their initial separation. 

\begin{figure}
\centering
\includegraphics[scale=1.0]{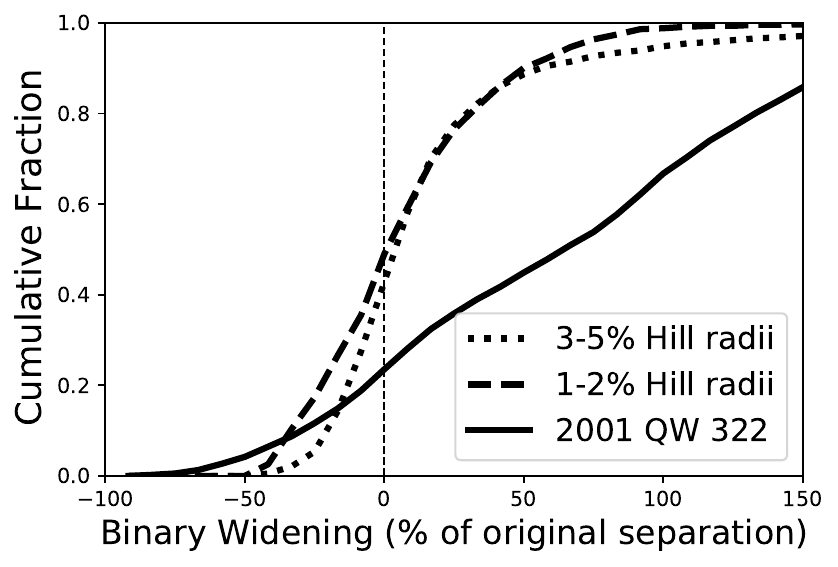}
\caption{{\bf Cumulative distribution of orbital separation change.} The net change in separations of three populations of binaries with different initial separations. Separation change is measured as a percentage of their initial separation with a negative widening meaning that the binary separation has actually decreased. A binary that has been lost over the 4 billion years of simulated time is assigned a separation change in accordance with its last recorded separation before disassociation.}
\end{figure} 

We should also note that Supplementary Figure 7 demonstrates that not all binaries widen over time. In the case of our 2001 QW$_{322}$ analogs, however, most (over 75\%) do. This is to be expected, since these are very ``soft binaries'' relative to the velocities of their perturbers \cite{heg75, bah85}. As initial binary separations become tighter (and therefore ``harder''), we see less of a bias toward semimajor axis expansion, and gradual tightening becomes more common. Among our binaries with initial semimajor axes of 3--5\% Hill radius, 58\% are widened and 42\% are tightened. For the tightest binary ensemble we model (1--2\% Hill radius), the widened fraction falls to 51\%. 

\bibliography{Cites.bib}

\end{document}